\documentclass[manuscript]{raa}           
\usepackage{graphicx,times}
\usepackage{natbib}
\usepackage{amssymb,amsmath}
\bibpunct{(}{)}{;}{a}{}{,}

\usepackage[a4paper=true,dvipdfm=true,pagebackref=true]{hyperref}
\hypersetup{pdftitle = The title of my PDF, pdfauthor = My name, pdfsubject= The subject, pdfkeywords = keyword1 keyword2 keyword3} 
\hypersetup{colorlinks = true, linkcolor = green, anchorcolor = red, citecolor = blue, filecolor = red, pagecolor = red, urlcolor = red}

\def\arcmin{\hbox{$^{\prime}$}}

\begin{document}

   \title{The Interstellar Medium: The Key Component in Galactic Evolution and Modern Cosmology}

 \volnopage{ {\bf 20XX} Vol.\ {\bf X} No. {\bf XX}, 000--000}
   \setcounter{page}{1}

   \author{Carl Heiles\inst{1}, Di Li\inst{2,3,4}, Naomi McClure-Griffiths\inst{5}, Lei Qian
      \inst{2,3},  Shu Liu\inst{2,3}
   }

   \institute{
        Department of Astronomy, University of California, Berkeley, 601 Campbell Hall 3411, Berkeley, CA 94720-3411; {\it heiles@astro.berkeley.edu} \\
         \and
	National Astronomical Observatories, Chinese Academy of Sciences, Beijing 100101, China\\
         \and
         Key Laboratory of FAST, National Astronomical Observatories, Chinese Academy of Sciences, Beijing 100101, China\\
         \and
         University of Chinese Academy of Sciences, Beijing 100049, China\\
	 \and
         Research School for Astronomy \& Astrophysics, Australian National University, Canberra, ACT 2611, Australia\\
\vs \no
   {\small Received 20XX Month Day; accepted 20XX Month Day}
}
\abstract{
The gases of the interstellar medium (ISM) possess orders of magnitude more mass than those of
all the stars combined and are thus the prime component of the baryonic universe. With L-band surface sensitivity even better than the planned phase one Square-Kilometer-Array (SKA1), the Five-hundred-meter Aperture Spherical radio Telescope (FAST) promises unprecedented insights into two of the primary components of ISM, namely, the atomic hydrogen (HI) and the hydroxyl molecule (OH). We discuss here the evolving landscape of our understanding of ISM, particularly, its complex phases, the magnetic fields within, the so-called dark molecular gas (DMG), high velocity clouds, and the connection between local and distant ISM. We lay out, in broad strokes, several expected FAST projects, including an all northern-sky high-resolution HI survey (22,000 deg$^2$, 3\arcmin\ FWHM beam, 0.2 km/s), targeted OH mapping, searching for absorption or
masing signals, and etc. Currently under commissioning, the commensal observing mode of FAST will be capable of simultaneously obtaining HI and pulsar data streams, making large-scale surveys in both science areas more efficient. 
}

   \authorrunning{C. Heiles et al. }            
   \titlerunning{The Interstellar Medium to be viewed by FAST}  
   \maketitle

%
\section{Introduction: Cosmological Evolution and the Interstellar Gas}           
\label{sect:intro}

Our knowledge of how the Universe has evolved is completely different
today from what it was two decades ago. From cosmology and extragalactic
astronomy we now have a theory based on cold, dark matter, called DM,
that explains with reasonable accuracy the Universe from its first few
minutes, through the formation of the first galaxies, to the present era
where clusters of galaxies dominate the Universe (e.g. \citealt{Sommer-Larsen+etal+2003}, \citealt{Komatsu+etal+2009}). This theory has been applied in
simulations of the Universe that show streams of gas coming together to
form the essential building blocks of the Universe, galaxies. These
galaxies in turn come alive through bursts of star formation, lose gas
to their surroundings and simultaneously accrete new gas to continue
their ravenous star formation habit. Eventually after billions of years
when their gas supplies run out they cease their star formation, living
on perpetually as increasingly red objects filled with low-mass
stars. 

Thus, the evolution of galaxies is inextricably linked to the
interstellar gas, and this is where the models break down. We know that
the life cycle of the Milky Way and most galaxies involves a constant
process of stars ejecting matter and energy into the interstellar mix,
from which new stars then condense, continuing the cycle.  For this part
of the theory observational or empirical assumptions are inserted,
rather than the detailed physics that drives the rest of the models.  If
we are to understand how galaxies evolve, we must first understand the
physics of their evolution in environments we can observe in detail. Our
own Galaxy, the Milky Way, provides us with the closest laboratory for
studying the evolution of gas in galaxies, including how galaxies
acquire fresh gas to fuel their continuing star formation, how they
circulate gas and how they turn warm, diffuse gas into molecular gas and
ultimately, stars. The Milky Way is a very complex
ecosystem. Just as ecosystems on Earth involve many elements linked
together by a common source of nutrients and energy flow, the Milky Way
ecosystem consists of stars fuelled by a shared pool of gas in the
interstellar medium (ISM) and energy that flows back and forth between
the stars, the ISM and out of the Galactic disk. 

We need to understand the details of these gasdynamical processes that
lie at the very heart of the astrophysics. How, exactly, do galaxies
form stars, acquire fresh gas, recycle their own gas to form new stars,
respond to and dissipate the large-scale energy inputs from supernova
and galactic shocks on scales from galactic to sub-parsec. These
overarching processes are, in turn, affected by the detailed physical
conditions. In particular, these include pressure, temperature,
ionization state, magnetic field, degree of turbulence, chemical
composition, morphology, and the effects of gravity.


\section{Interstellar Matter in the Milky Way}
\label{sect:milky}

Once thought to be a simple, quiescent medium, the ISM is now known to
include a number of diverse constituents, which exhibit temperatures and
densities that range over six orders of magnitude. The ISM is composed
of gas in all its phases (ionized, atomic and molecular), dust, high
energy particles and magnetic fields, all of which interact with the
stars and gravitational potential of a galaxy to produce an
extraordinary, dynamic medium.

\subsection{Phases of the ISM}

\subsubsection{The Classical Five Phases}

In the Solar vicinity, astronomers generally recognize five phases of
interstellar gas: dense molecular clouds, which are traced by CO line
emission; the atomic Cold and Warm Neutral Media (CNM and WNM), traced
by the 21-cm line; the Warm Ionized Medium (WIM), traced by pulsar
dispersion and H$\alpha$ emission; and the Hot Ionized Medium (HIM),
traced by X-ray emission. The atomic and molecular phases have
comparable mass (the molecular phase rises and dominates toward the
Galactic interior's `Great Molecular Ring') and the WIM is somewhat
less.

\subsubsection{Dark Gas: The Sixth Phase}

However, there lurks a sixth phase: Dark Molecular Gas, in which
Hydrogen is molecular but the usual H$_2$ tracer, CO emission, is
absent. Dark Molecular Gas was discovered (we believe) when 
\citet{Dickey+etal+1981} found OH in absorption against high
Galactic latitude continuum sources. Important and extensive
confirmatory absorption measurements by \citet{Liszt+Lucas+1996} 
and \citet{Lucas+Liszt+1996} found that OH and HCO$^+$ are commonly observed against
such sources. These two molecules are much more easily seen in
absorption than emission because their excitation temperatures $T_x$ are
low. In the relatively thin clouds where they reside the
collisional excitation rates are small so that $T_x \ll T_k$, which was borne out by our
analysis (\citealt{Li+etal+2018a}, also Sec. \ref{darkmol} in this work) of the Millennium survey data \citep{Heiles+Troland+2003}.

Historically, molecular lines were seen mainly in emission towards
the standard dense molecular clouds, and CO was emphasized to the extent
that its presence {\it defined} molecular gas. While most astronomers remain
unaware that Dark Molecular Gas is so prominent,  a few courageous radio
astronomers have pursued Dark Molecular Gas through spectroscopy. Liszt
and his collaborators, primarily Lucas and Pety, have observed
absorption and emission lines of OH, HCO$^+$, and CO to establish
abundance ratios, and they have mapped CO in the Dark Molecular Gas
regions; this work currently culminates in the comprehensive
presentation of \citet{Liszt+Pety+2012}, who show CO emission maps together
with CO, HCO$^+$, and OH absorption spectra for 11 continuum
sources. \citet{Cotten+etal+2012} and \citet{Cotten+Magnani+2013} mapped CH
and OH around the dense molecular cloud MBM40. \citet{Allen+etal+2012, Allen+etal+2015} 
found extensive OH emission in their map; at most positions, CO emission was absent.  
Studies of individual clouds, e.g.\ the Taurus Molecular Cloud (TMC, \citealt{Xu+etal+2016}, \citealt{Xu+Li+2016}), 
also tend to find substantial CO-dark molecular gas.
Quite generally, observers find the mass of Dark Molecular Gas to be comparable to that
of the CO-bright molecular gas.

It's not just radio astronomers! \citet{Grenier+etal+2005} 
used the Energetic Gamma Ray Experiment Telescope (EGRET) to map the
diffuse Galactic gamma rays produced by the interaction of cosmic rays
with H-nuclei. The gamma-ray intensity is proportional to the total
H-nuclei column density, whether in atomic or molecular form; comparing
with CO emission unveils the Dark Molecular Gas (DMG). They find that the Dark
Molecular Gas is very common throughout the Galaxy, even in the
interior. It surrounds all the nearby CO clouds and bridges the dense
cores with the broader atomic clouds, thus providing a key link in the
evolution of interstellar clouds. The general trend of the fraction of the gamma-ray identified DMG are found to
follow those of simple hydrides \citep{Remy+etal+2018}. As they conclude, ``The relation
between the masses in the molecular, dark, and atomic phases in the
local clouds implies a dark gas mass in the Milky Way comparable to the
molecular one.''

\subsection{The CNM}

\subsubsection{The CNM and Its Relation to Dark Molecular Gas}

It seems almost certain that Dark Molecular gas is a transition state
between the CNM and classical molecular clouds.  This emphasizes the
importance of studying the CNM together with the prime DMG tracers, OH
and HCO$^+$. They will tell us where the formation of molecules is
initiated, and the detailed comparison of the atomic and molecular
spectral lines will provide the temperature and density. Moreover, we
expect the details of the DMG transition region to depend not only on
physical conditions but also cloud {\it morphology}. Morphology
determines whether UV photons can penetrate to destroy molecules via
photodissociation or photoionization. It seems to us very unlikely that
one can understand the transition between atomic and molecular gas
without understanding the effect of UV photons, and thus cloud
morphology. Moreover, there are hints that cloud morphology is affected
by the magnetic field; after all, magnetic forces are one of the
important forces on the ISM (the others being turbulent pressure, cosmic
ray pressure (coupled to the gas by the magnetic field), thermal
pressure, and gravity).

\subsubsection {CNM: Physical conditions}

Our current knowledge of physical conditions and morphology in the CNM
depends overwhelmingly on results from the Millennium survey of \citet{Heiles+Troland+2005} (HT), who used Arecibo with long integration times
suitable for detecting Zeeman splitting. For the HI line in absorption,
HT derived column densities, temperatures, turbulent Mach number, and
magnetic fields.  HI CNM Column densities are usually below $10^{20}$
cm$^{-2}$ with a median value $N(HI)_{20} \sim 0.5$. The median spin
temperature $T_s \sim 50$ K and the median turbulent Mach number $\sim
3.7$. The median magnetic field $\sim 6$ $\mu$G \citep{Heiles+Troland+2005}; this is a
statistical result and individual detections are too sparse to make a
meaningful histogram.

\subsubsection{CNM: Morphology}

Regarding morphology, \citet{Heiles+Troland+2003}, in their \S\ 8, present one of the few
discussions of 3-d CNM morphology. By assuming reasonable values for the
thermal gas pressure and comparing observed column densities, shapes and
angular sizes as seen on the sky, they find that interstellar CNM
structures cannot be characterized as isotropic. The major argument is
that a reasonable interstellar pressure, combined with the measured
kinetic temperature, determines the volume density; this, combined with
the observed column density, determines the thickness of the cloud along
the line of sight. This dimension is almost always much smaller than the
linear sizes inferred from the angular sizes seen on the sky. \citet{Heiles+Troland+2003}
characterize the typical structures as `blobby sheets', and this applies
for angular scales of arcseconds to degrees. An alternative, which they
did not discuss, is that the structures are more spherical, but spongelike inside. 
(If they are spongelike, what fills the holes?) 

Apart from this general argument, only a very few individual
interstellar structures have been characterized morphologically.  One
important reason for the small numbers is the difficulty of mapping the
21-cm line with simultaneously high brightness temperature sensitivity and high angular
resolution. Single dishes have high sensitivity and low resolution,
while interferometers have high resolution but low sensitivity. The
GALFA 21-cm line survey \citep{Peek+etal+2011a}, which is a fully-sampled
survey of the entire Arecibo sky (declination $0 ^\circ$ to $39^\circ$,
about 1/3 of the entire sky) provides the best of both worlds, with
angular resolution 3.4 arcminutes and sensitivity 0.1 K. FAST will do
even better!

\begin{figure}[h!]
\vspace{-4ex}
\begin{center}
\leavevmode
\includegraphics[scale=.5]{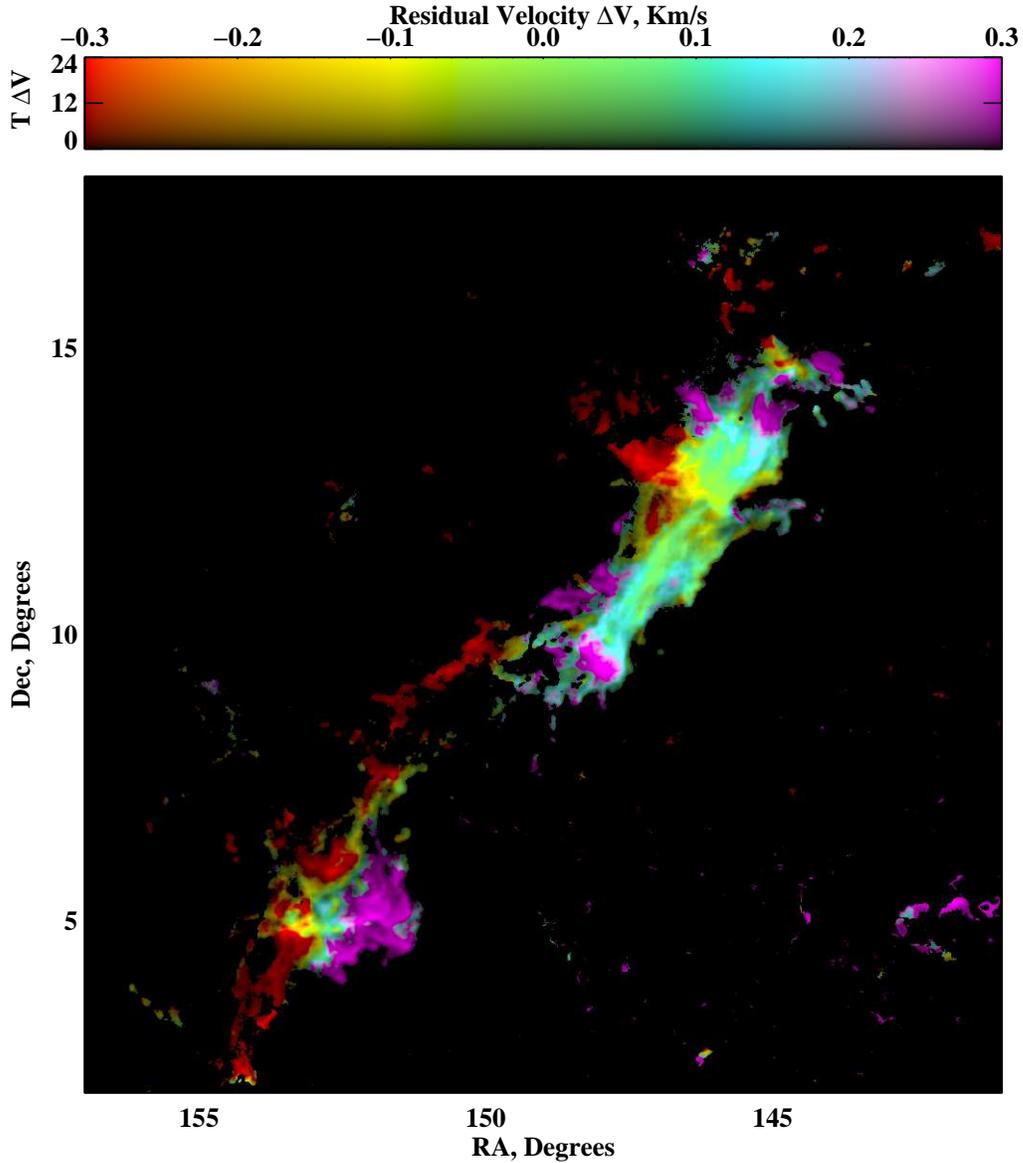}
\end{center}
\vspace{-10ex}
\caption{\footnotesize Image of CNM (from GALFA data) in the Local Leo
  Cold Cloud (LLCC). Color indicates residual velocity (after
  subtracting a velocity gradient). \label{leo}}
\vspace{-2ex}
\end{figure}

The most spectacular result is the `Local Leo Cold Cloud' (LLCC: \citealt{Peek+etal+2011b}, \citealt{Meyer+etal+2012}), which we characterize as a remarkably
thin sheet. Figure \ref{leo} images the LLCC, with color indicating the
residual radial velocity after subtracting a clear velocity gradient
along the length of the cloud, which amounts to about 1 km s$^{-1}$. The
cloud temperature, in some places less than 20 K, is measured at a few
positions both by emission/absorption of the 21-cm line and by
optical/UV spectroscopy. The UV absorption lines of CI provide the
interstellar pressure, $P/k \sim 60000$ cm$^{-3}$ K, which exceeds the
typical interstellar pressure by well over an order of magnitude. The
pressure and temperature provide the density, $n(HI) \sim 3000$
cm$^{-3}$; the observed column density, $N(HI)_{20} \sim 0.04$, provides
the thickness along the line of sight---about 200 AU. The cloud distance
lies between 11 and 24 pc; this close distance, combined with the
absorption of diffuse X-rays mapped by ROSAT, confirms the absence of
hot gas inside the Local Bubble. With the LLCC's angular dimension of
several degrees, its extent on the plane-of-the-sky is about 1 pc. The
aspect ratio (thickness/extent) is about $10^{-3}$! With
a thickness of only 200 AU, the time scale for changing along the line
of sight is only a few thousand years---within recorded human history!
The $\sim 10$-year time scale for variability of the Na optical
absorption line against stars is satisfying confirmation of this rapid
evolution \citep{Meyer+etal+2012}.

\subsection {OH and HCO$^+$: Tracers of Dark Molecular Gas} \label{darkmol}

In Dark Molecular Gas, 
Hydrogen is molecular but the usual H$_2$ tracer, CO emission, is largely absent. CO
requires protection from UV radiation, which occurs either from CO
self-shielding or dust extinction. In these cases, H$_2$ is well-traced by
OH (\citealt{Lucas+Liszt+1996}, \citealt{Liszt+Lucas+2004}) and also by HCO$^+$ (\citealt{Liszt+Pety+2012}, \citealt{Liszt+etal+2010}), so much so that the observed
column densities are linearly related:

\begin{equation} \label{ohtoh2}
{N(OH) \over 2N(H_2)} = 0.5 \times 10^{-7} \ \ \ \ ; \ \ \ \ {N(HCO^+) \over 2N(H_2)} = 1.2 \times 10^{-9}
\end{equation}


\noindent and, of course, this means OH and HCO$^+$ are also linearly
related, with 

\begin{equation} \label{combo}
{N(HCO^+) \over N(OH)}  = 0.03
\end{equation}

These linear relationships can be understood as the result of
ion-molecule reactions arising in cool regions where Carbon is C$^+$. The
relevant reaction chain for OH production involves 7 rapid ion-molecule
reactions involving e$^-$, H, O, OH$^+$, OH$_2^+$, H$_2$, H$_2^+$, H$_3^+$, and
H$_3$O$^+$ \citep{Wannier+etal+1993}. Having formed OH, we obtain HCO$^+$ \citep{Lucas+Liszt+1996}:

\begin{equation}
{\rm C^+  +  OH \rightarrow CO^+  +  H}  \ \ \ \ ; \ \ \ \ {\rm CO^+  +  H_2 \rightarrow HCO^+ +  H} 
\end{equation}

\noindent The only problem with this scheme is that the ratio
$N(HCO^+)/N(OH)$ is predicted to be about 20 times smaller than the
observed ratio, which is a sad reflection on our understanding of
astrochemistry. Nevertheless, the observed ratios for OH, HCO$^+$, and
H$_2$ in equations \ref{ohtoh2} and \ref{combo} are very
robust, which is an empirical demonstration that OH and HCO$^+$ are
excellent tracers of H$_2$, particularly at low column densities where
CO cannot survive.

\subsubsection{The Millennium Survey's OH as a Tracer of Dark Molecular Gas}

\begin{figure}[h!]
\begin{center}
\leavevmode
\includegraphics[scale=.75]{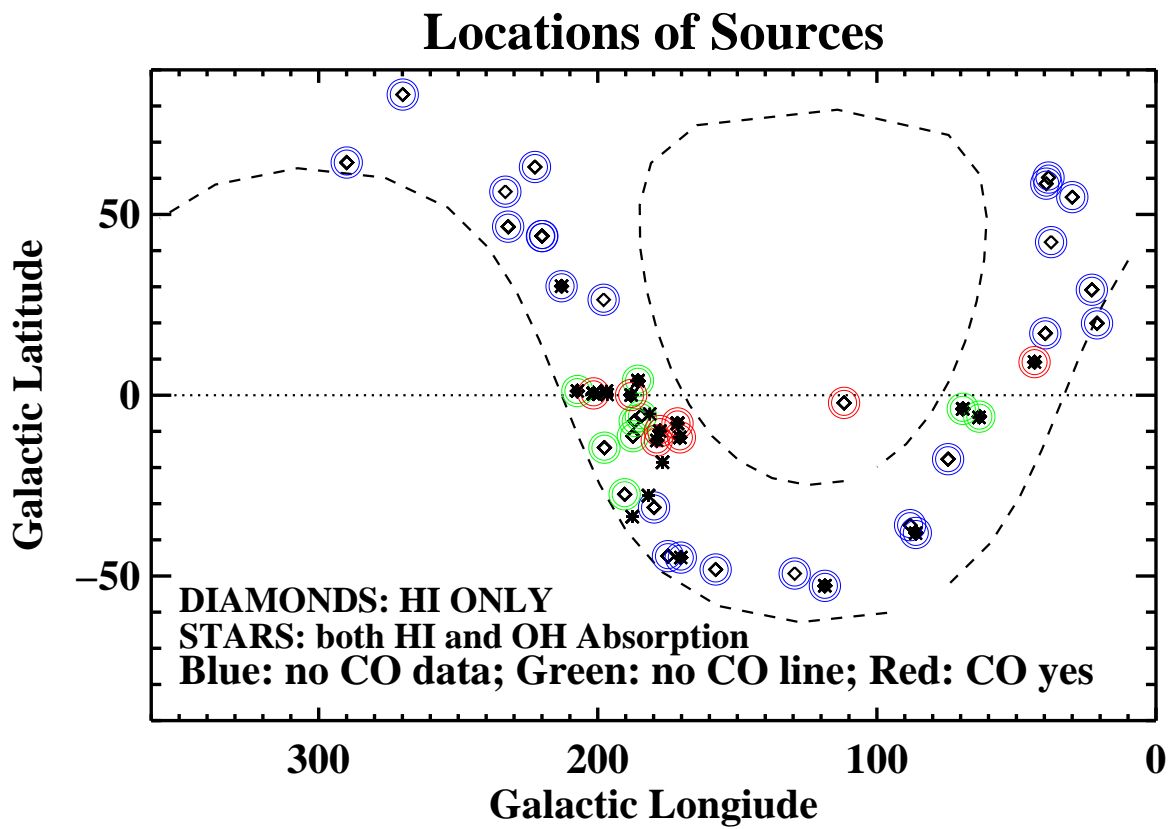}
\end{center}
\vspace{-7ex}
\caption{\footnotesize Map of sources observed in the Millennium survey. Diamonds are
  sources showing HI absorption only. Stars show both HI and OH
  absorption. Blue circles show sources having no CO data. Green
  circles show sources having CO data, but no detected CO line. Red
  circles show sources having detected CO line. \label{tst4}}
\end{figure}

In the HT Millennium survey there was `extra' spectrometer capability
that allowed simultaneous observation of the two `main' lines (1665 and
1667 MHz) of ground-state OH. The long integration times required for
detecting HI Zeeman splitting provided excellent sensitivity for OH in
absorption.  Figure \ref{tst4}, which is based on those Millennium
survey data,
shows that OH traces Dark Molecular gas. Black stars show sources with
detected OH absorption. Red circles show sources with detected CO
emission\footnote{Here, we used CO emission data from Dame's compilation
  ({\tt http://www.cfa.harvard.edu/rtdc/CO/}). Other surveys are \citet{Liszt+1994} and \citet{Liszt+Wilson+1993}, which we have not yet
  examined. However, for one source, 3C207, Dame found emission and
  \citet{Liszt+1994} did not. This demonstrates the need for better data!}
and green circles show no CO emission. Therefore, sources with both
black stars and green circles have OH absorption but no CO emission, so
these show Dark Molecular Gas. (Blue circles have no CO data).

There are eight sources with OH absorption but no CO emission. There are
7 sources showing both OH absorption and CO emission. Thus {\it Dark
  Molecular Gas (with no CO) is as common as Undark Molecular Gas (with
  CO).}  This result is consistent with \citet{Lucas+Liszt+1996}, who
observed 30 mm-wave continuum sources for absorption of HCO$^+$ and of
CO. HCO$^+$ absorption lines are common, while CO absorption lines are
uncommon.

\subsubsection{OH as a Tracer: Absorption {\it vs.} Emission}

\begin{figure}[h!]
\begin{center}
\leavevmode
\includegraphics[scale=.24]{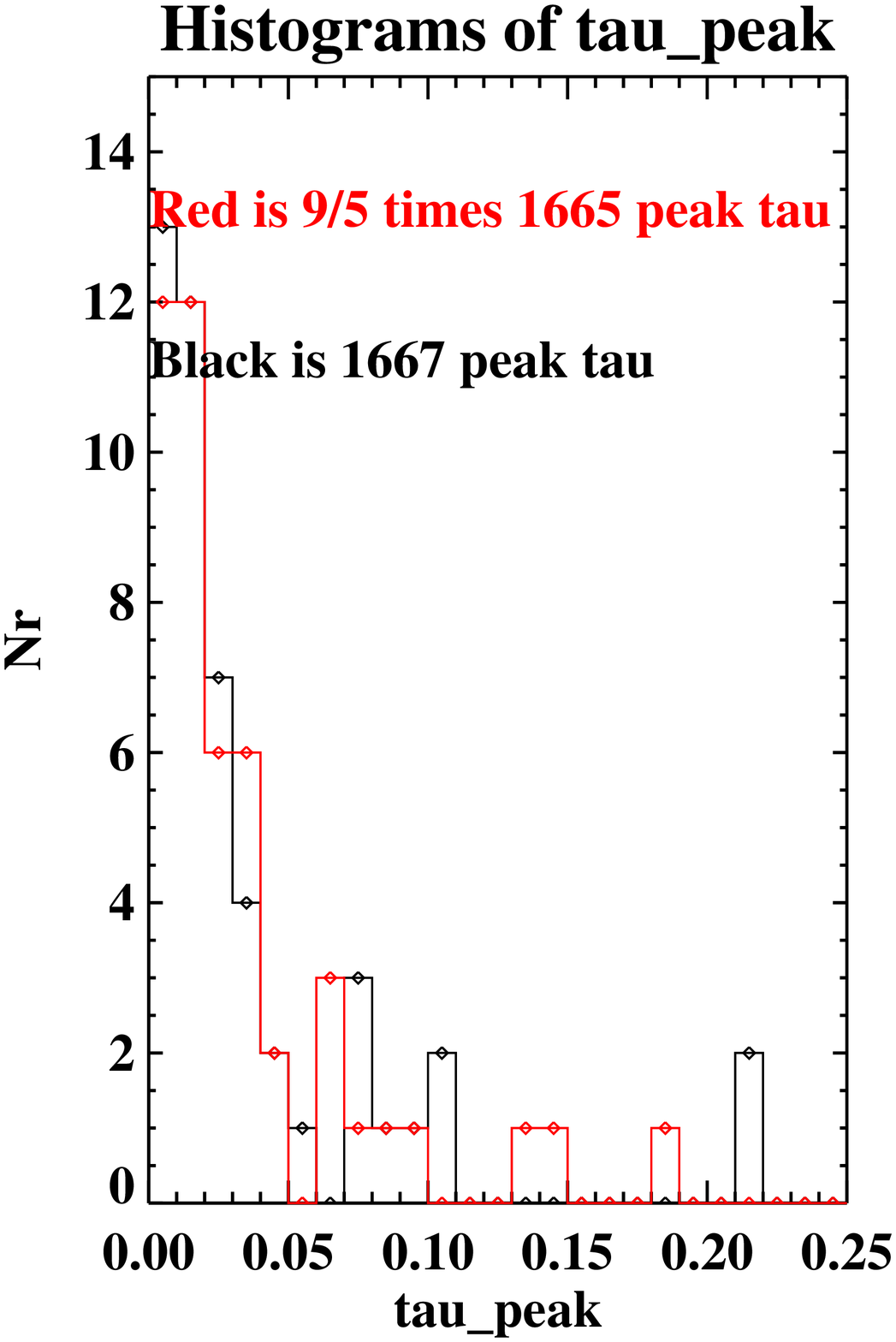}
\includegraphics[scale=.24]{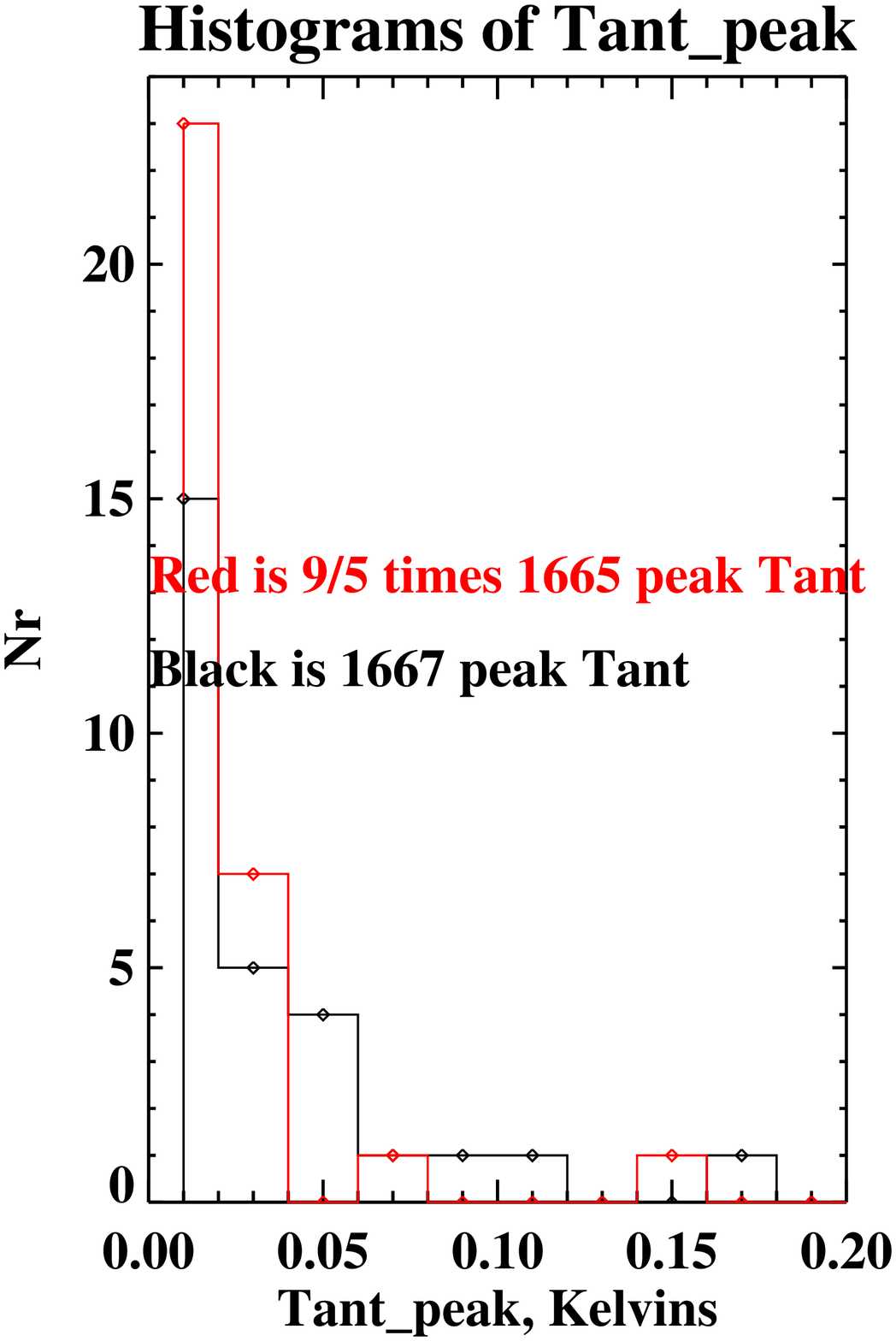}
\includegraphics[scale=.24]{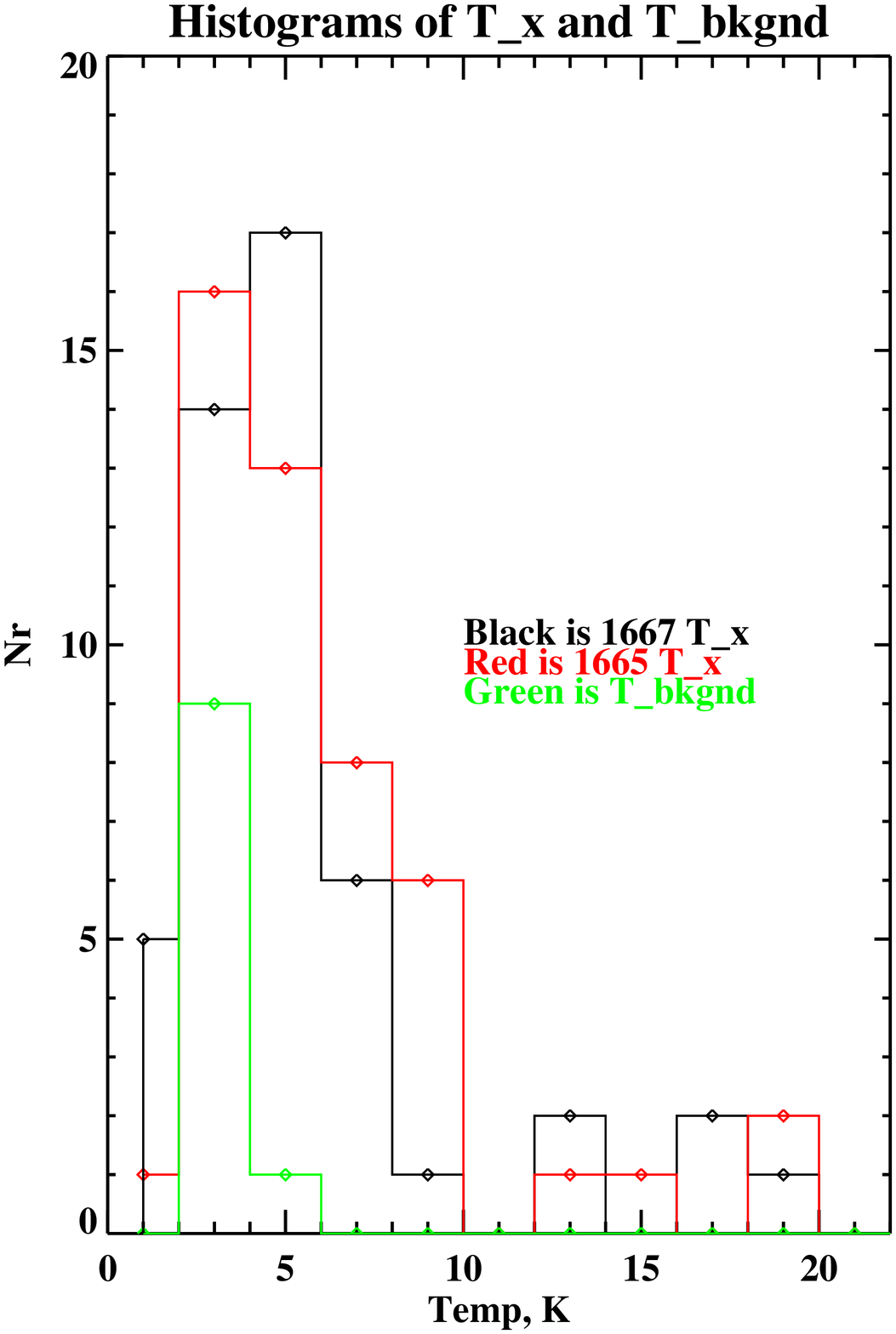}
\end{center}
\vspace{-5ex}
\caption{\footnotesize Histograms of peak optical depth, antenna temperature in
  emission, and excitation temperature for the Gaussian
  components. Black shows 1667 MHz and red shows 9/5 times the peak optical
  depth for 1665 MHz.  \label{tst3_taus}}
\vspace{-2ex}
\end{figure}

Figure \ref{tst3_taus} exhibits as-yet unpublished histograms of the
Millennium survey OH optical depths $\tau$, emission line antenna
temperature $T_A$, and excitation temperatures $T_x$. We decomposed all
OH lines into Gaussian components and made histograms of the peaks. For
the 1665 MHz line, the optical depths and antenna temperatures are
multiplied by 9/5 so that the scales for the two lines are identical if
the excitation is thermal. The histograms for the two lines do, indeed,
look similar, so neither line is anomalously excited with respect to the
other. This is reflected in the right panel of Figure \ref{tst3_taus},
which shows the histogram of excitation temperatures for the Gaussian
components. Typically, $T_x \sim 5$ K. The green histogram represents
the background continuum brightness temperature $T_C$, which consists of
the CBR plus the Galactic synchrotron background; typically, for these
sources that lie away from the Galactic plane, $T_C \sim 3.5$ K.
 
The Millennium survey continuum sources generally have flux density $S
\gtrsim 2$ Jy, so they produce continuum antenna temperatures in excess
of about 20 K at Arecibo.  For our weakest optical depths $\tau = 0.01$,
the observed absorption line deflection is $\gtrsim 0.2$ K. In contrast,
the OH emission lines are much weaker; nearly all the emission lines
have deflections $\lesssim 0.1$ K. Thus, OH is much easier to detect in
absorption than in emission.  This is easy to understand when
considering the excitation temperature. For a frequency-switched
emission spectrum of a single OH feature having peak optical depth
$\tau$ seen against a continuum background brightness temperature $T_C$,
the observed antenna temperature $T_A$ is

\begin{equation} \label{diff}
\Delta T_A = \left[T_x - T_C\right] \left[1 - \exp(-\tau)\right] \ \ 
\end{equation}

\noindent Because $T_A \propto (T_x - T_C)$, the emission line
intensities are significantly reduced. 

\section{GALACTIC EVOLUTION AND THE INTERSTELLAR MEDIUM}

\subsection{How Does Energy Flow in the Disk and Between the Disk and
  Halo?}  

The Milky Way is not a closed system. The evolution of the
Milky Way is significantly impacted by the two-way flow of gas and
energy between the Galactic disk, halo, and intergalactic medium. We
have long known that the atomic hydrogen halo extends far beyond the
disk of the Galaxy. In recent years we have come to realize that the
halo is also a highly structured and dynamic component of the
Galaxy. Although we can now detect hundreds of clumped clouds in the
atomic medium of the halo \citep{Ford+etal+2010}, we
are far from understanding the halo's origin and its interaction with
the disk of the Galaxy. 

It has been proposed that there may be two
dominant sources of structure in the halo: one is the outflow of gas
from the Galactic disk, and the second is the infall of gas from
extragalactic space. The relative importance of these sources and their
effects on the global evolution of the Milky Way are not known.  

It seems that a significant fraction of the structure of gas in the
Galactic halo may be attributed to the outflow of structures formed in
the disk, but extending into the halo. An example of such a structure
may be an HI supershell. There are several examples of HI supershells
that have grown large enough to effectively outgrow the Galactic HI disk
(e.g. \citealt{McClure-Griffiths+etal+2003}, \citealt{McClure-Griffiths+etal+2006}). 
When this happens, the rapidly decreasing density of the Galactic
halo does not provide sufficient resistance to the shell's expansion and
it will expand unimpeded into the Galactic halo, creating a chimney from 
disk to halo. These chimneys supply hot, metal-enriched gas to the
Galactic halo and may act as a mechanism for spreading metals across the
disk. It has been theorized that HI chimneys in the disk of the Galaxy
may be a dominant source of structure for the halo through a Galactic
Fountain model (\citealt{Shapiro+Field+1976}, \citealt{Bregman+1980}). Some Fountain
models, e.g. \citet{deAvillez+2000}, predict that cold cloudlets should
develop out of the hot gas expelled by chimneys on timescales of tens of
millions of years. Other Fountain theories, e.g. \citet{MacLow+etal+1989}, 
suggest that the cool caps of an HI supershell will
extend to large heights above the Galactic plane before they break. Once
broken, the remains of the shell caps could be an alternate source of
small clouds for the lower halo.

Recent observational work has placed this theory on firmer footing,
showing that the clumped clouds that populate the lower halo are not
only more prevalent in regions of the Galaxy showing massive star
formation than in less active regions, but that they extend higher into
the halo in these regions \citep{Ford+etal+2010}. 
Now we would like to know how these clouds survive as cold,
compact entities as they traverse the hot lower halo, with temperatures
up to $10^5$ K, and whether the cloud structures can tell us anything about
their past journey through the halo. 

Our ability to answer these questions is significantly hampered by our
inability to image the detailed physical structure of the HI in the
halo.  With the present generation of all-sky surveys at 15\arcmin\ angular
resolution and 1 km s$^{-1}$ spectral resolution, it is not possible to
explore the thermal and physical structure of these cloudlets, which may
contain information about their origin, motion and evolution. Follow-up
observations on several individual cloudlets show a tantalizing
glimpse of possible head-tail structure and other evidence of
interaction. 

These observations may offer the opportunity to resolve a directional
ambiguity with these clouds. In general we can measure only the absolute
value of the z component of the cloud velocity with respect to the
Galactic Plane, i.e.\ we cannot determine whether the clouds are moving
towards or away from the Galactic Plane. Finding head-tail structure in
the clouds can help resolve this ambiguity by showing the direction of
motion. At present these follow-up observations are prohibitive,
requiring hundreds of hours with interferometers to reach 100 mK at
arcminute scales. The sensitivity and spectral resolution of FAST will
allow us to study the spatial and spectral structure of disk-halo clouds.
We believe that the structure of the lower halo is related to the star
formation rate in the disk below. FAST has greater sky coverage than
Arecibo and extends into regions of the Galaxy where the star formation
rate is of interest. 

\subsection{High Velocity HI Associated with the Milky Way and
Magellanic System}

Cosmological simulations predict that gas accretion onto galaxies is
ongoing at $z = 0$. The fresh gas is expected to provide fuel for star
formation in galaxy disks \citep{Maller+Bullock+2004}. In fact, galaxies
like the Milky Way must have received fresh star formation fuel almost
continuously since their formation in order to sustain their star
formation rates. High velocity clouds (HVCs), first identified in HI 21
cm emission at anomalous (non-Galactic) velocities, have been suggested
as a source of fuel \citep{Quilis+Moore+2001}. Some of the HI we see in
the halo of the MW comes from satellite galaxies, some is former disk
material that is raining back down as a galactic fountain, and some may
be condensing from the hot halo gas \citep{Putman+2006}. The relative
fractions of these structures are unknown. Furthermore, the detailed
physics of how gas comes into the Milky Way disk is still unknown. How
much gas flows into the disk through the halo, how fast does it flow,
and what forces act on it along the way?  Is the accretion of the
Magellanic Stream a template for galaxy fuelling?  

The Magellanic Stream provides the closest example of galaxy
fuelling. The Magellanic Stream, which extends almost entirely around
the Milky Way \citep{Nidever+etal+2010}, is gas stripped off the nearby
Small Magellanic Cloud during its interaction with the Large Magellanic
Cloud and the Milky Way. While the Magellanic Leading Arm is believed to
be closely interacting with the Milky Way disk, the Northern tip of the
Magellanic Stream is furthest from the Milky Way and contains a wealth
of small scale structure \citep{Stanimirovic+etal+2008}. By studying the
details of the physical and thermal structure of the Magellanic Stream
and its interaction with the Milky Way we will reveal its origin and
evolution.  

\subsection{High and Intermediate Velocity Clouds in the Milky Way}

High and intermediate velocity clouds (HVCs and IVCs) are believed to
play important roles in both the formation and the evolution of the
Galaxy. Some HVCs may be related to the Galactic Fountain; some are
tidal debris connected to the Magellanic Stream (\citealt{Putman+etal+2003}, 
\citealt{Stanimirovic+etal+2008}) or other satellites; some may be infalling
intergalactic gas, and some may be associated with dark matter halos and
be the remnants of the formation of the Local Group. The structure and
distribution of high velocity gas probes tidal streams and the building
blocks of galaxies providing critical information on the evolution of
the Milky Way system. One problem that has long plagued HVC studies is
the almost complete lack of distances to these objects, which has
traditionally sparked debates about whether HVCs are Galactic or
extragalactic.  Although the distance problem is improving somewhat with
increased number of distances using absorption towards halo stars, the
problem remains. Regardless, the consensus now is that there are
probably a variety of HVC origins.

\begin{figure}[h!]
\begin{center}
\leavevmode
\includegraphics[scale=.8, bb=145 120 500 694, clip]{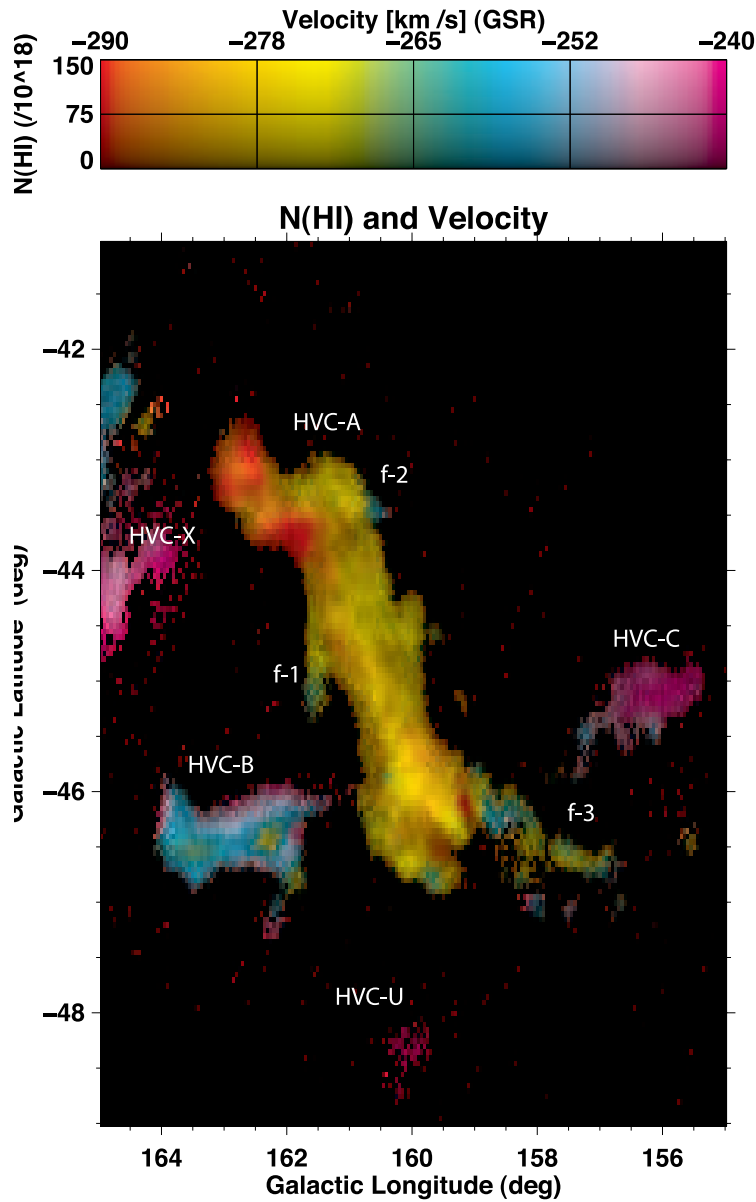}
\end{center}
\vspace{-4ex}
\caption{\footnotesize The shards of \citet{Peek+etal+2007}'s HVC. HI column
  density (brightness) and central velocity (color).  \label{peek}}
\end{figure}

The nearby HVCs can be used as probes of the thermal and
density structure of the Galactic halo. HVCs have, in general, a high
velocity relative to their ambient medium, which results in a ram
pressure interaction between the cloud and medium. Recent HI
observations have indeed shown that a significant fraction of the HVC
population have head-tail or bow-shock structure \citep{Bruns+etal+2000}. 
By comparing the observed structures and thermal distributions
within them to numerical simulations it is possible to determine basic
physical parameters of the Galactic halo such as density, pressure, and
temperature. In situations where the density and pressure of the ambient
medium are known by independent methods the problem can be turned around
to determine the distance to the observed clouds (\citealt{Peek+etal+2007}; 
Figure \ref{peek}). Recent large-scale single dish surveys, the Galactic All-Sky Survey
(GASS; \citealt{McClure-Griffiths+etal+2009}, \citealt{Kalberla+etal+2010}) and the
HI-Galactic Arecibo L-Feed Array (GALFA; \citealt{Peek+etal+2011a}) have
produced outstandingly detailed and sensitive images of HI associated
with the Milky Way and Magellanic Stream. These surveys offer high
spectral resolution, revealing how the spectral structure of disk-halo
structures give important clues to their evolution. With these
improvements we are beginning to resolve the structure of some of the
mid-sized HVCs. 

However, we do not yet have the sensitivity or resolution necessary to
study the detailed physical and thermal structure over a large number of
HVCs. It seems that the structure that we do see is just the tip of the
iceberg.  GASS is also revealing a wealth of tenuous filaments
connecting HVCs to the Galactic Plane at column densities of $N(HI) \sim
10^{18}$ cm$^{-2}$. As an example, Complex L shows that the HVCs are
interconnected by a thin, low column density filament and that this
filament appears to extend to the Galactic disk. Filaments of HVCs such
as these may be related to the “cosmic web” predicted in cosmological
simulations. Observations of the M31-M33 system (\citealt{Braun+Thilker+2004}, 
\citealt{Wolfe+etal+2013}) have revealed the local analogue of the cosmic web,
showing that very low column density gas ($N(HI) \sim 10^{16} - 10^{17}$
cm$^{-2}$) connects the two galaxies.

\section{The ISM at Moderate to High Redshift}

We want to learn about galaxy evolution in the cosmological context, so
we should pursue measurements at moderate to high redshift to the extent
possible. Owing to sensitivity requirements, from a practical standpoint
these are available in absorption, and in rare cases possibly in maser
emission from OH, H$_2$O, or CH$_3$OH. Highly redshifted 21-cm HI
absorption lines are seen in damped Ly$\alpha$ lines and also against
the occasional radio-bright continuum source \citep{Curran+etal+2011}; in
some of these cases, molecules also appear in absorption, especially OH
\citep{Chengalur+etal+1999}. Studying such systems will provide limited,
but unique, information on temperatures, chemical concentration, and
magnetic field strengths---and how these quantities change with
redshift.

\section{The Role of FAST}


\subsection{Expand the GALFA Survey to an All-Sky FAST-HI Survey}

We need to map the 21-cm line over the entire 22,000 deg$^2$ FAST
sky---i.e., we need to do with FAST what GALFA did with Arecibo to
obtain more sky coverage with the best combination of angular resolution
and sensitivity. This will map the ISM morphology generally, including
HVCs, and will enhance the sample of the unique structures discovered
with GALFA: compact clouds and fibers. This survey should use the
multibeam feed for observing efficiency and should cover at least
GALFA's velocity range ($\pm 800$ km s$^{-1}$) and resolution (0.2 km s$
^{-1}$). With a resolution of 2.9 arcmin and full sampling (pixel size $<$ 
1.5 arcmin) and 12 seconds per pixel, the survey requires about 40000
hours with a single-pixel feed. With the 19-element feed in drift-scan mode, 
this is a ~5000 hour project.

\subsection{Expand the Millennium Survey: HI Zeeman Splitting with
  Simultaneous OH Emission/Absorption}

\begin{figure}[h!]
\vspace{-4ex}
\begin{center}
\leavevmode
\includegraphics[scale=.7]{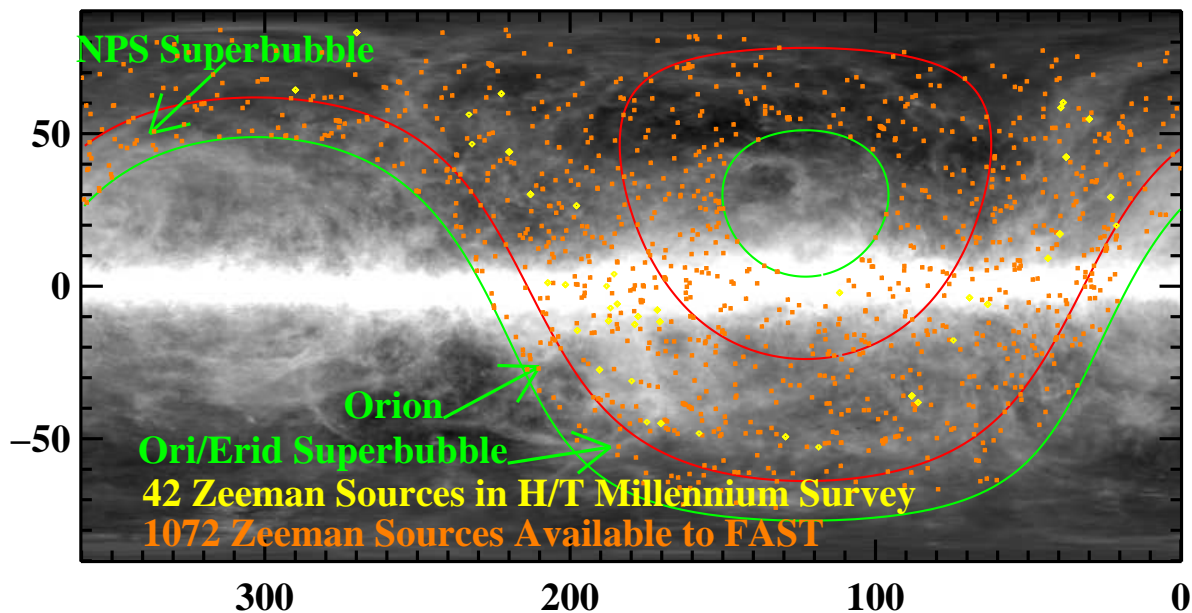}
\end{center}
\vspace{-2ex}
\caption{\footnotesize Gray-scale image of the velocity-integrated HI
  emission for the whole sky. Red and green lines enclose the Arecibo
  sky and FAST coverage regions. Orange dots show all radio continuum
  sources with fluxes exceeding 0.5 Jy, which makes them suitable for
  measuring Zeeman splitting. \label{millennium}}
\end{figure}

Figure \ref{millennium} says it all. The Millennium survey observed HI
Zeeman splitting and OH absorption and it covered 76 sources, of which
only 42 had detected magnetic fields. The statistical sample of measured
Zeeman splittings is woefully small and desperately needs to be
expanded. With FAST, a 0.5 Jy source has an antenna temperature roughly
equal to the system temperature, so the continuum is strong and Zeeman
sensitivity is high. There are 1072 such sources available to Fast, and
ultimately each one should be included in a new Millennium survey.

Zeeman-splitting observations are time-consuming: typically tens of
hours are required for each source. That works out to about 20,000 hours
of telescope time; this survey will keep FAST occupied for a long time! 

Surveying OH absorption is important for studying Dark Gas, and the OH
lines are best studied in absorption. Even so, they are weak. Clearly,
HI Zeeman-splitting and OH should be observed simultaneously because
they have similar sensitivity requirements. 

\subsection{Map OH emission in well-chosen regions}

Ultimately, understanding Dark Molecular Gas will require mapping of HI
and also the DMG tracers. The two best DMG tracers are OH and HCO$^+$;
of these, OH has a much smaller critical density and is consequently a
much better tracer in emission. FAST is the ideal telescope for this
purpose. Mapping can begin with those fields already mapped in CO by
\citet{Liszt+Pety+2012}, and continue with other CO fields, including the
peripheries of molecular clouds (e.g. \citealt{Cotten+Magnani+2013}, \citealt{Allen+etal+2012}).

\subsection{Moderate to High Redshift}

DLAs and other cosmologically relevant lines, either HI in absorption or
molecules---mainly OH---in absorption or maser emission, should be
systematically observed as they are found. 

The FAST's large diameter, filled aperture, and sky coverage provide the angular resolution and sensitivity required to make it the best telescope for HI and OH emission maps. Together with Arecibo, FAST is uniquely suitable in providing the sensitivity required for the very weak OH lines in emission/absorption, and it will also do an excellent job for HI. The sky coverage will provide a greatly expanded set of interstellar structures, which is essential for obtaining a reliable statistical sample. The proposed novel commensal survey mode (Commensal Radio Astronomy FAST Survey -- CRAFTS, \citealt{Li+etal+2018b}) will realize large-scale HI imaging simultaneously with pulsar search, making both types of surveys more efficient. We look forward to getting started!

\normalem
\begin{acknowledgements}
This work is supported by National Key R\&D Program of China No. 2017YFA0402600, the CAS International Partnership Program No.114A11KYSB20160008, and NSFC No. 11725313.

\end{acknowledgements}
  
\bibliographystyle{raa}
\bibliography{bibtex}

\begin{thebibliography}{47}
\providecommand\natexlab[1]{#1}
\providecommand\JournalTitle[1]{#1}

\bibitem[{Allen} {et~al.}(2015)]{Allen+etal+2015}
{Allen}, R.~J., {Hogg}, D.~E., \& {Engelke}, P.~D. 2015, \aj, 149, 123

\bibitem[{Allen} {et~al.}(2012)]{Allen+etal+2012}
{Allen}, R.~J., {Ivette Rodr{\'{\i}}guez}, M., {Black}, J.~H., \& {Booth},
  R.~S. 2012, \aj, 143, 97

\bibitem[{Braun} \& {Thilker}(2004)]{Braun+Thilker+2004}
{Braun}, R., \& {Thilker}, D.~A. 2004, \aap, 417, 421

\bibitem[{Bregman}(1980)]{Bregman+1980}
{Bregman}, J.~N. 1980, \apj, 236, 577

\bibitem[{Br{\"u}ns} {et~al.}(2000)]{Bruns+etal+2000}
{Br{\"u}ns}, C., {Kerp}, J., {Kalberla}, P.~M.~W., \& {Mebold}, U. 2000, \aap,
  357, 120

\bibitem[{Chengalur} {et~al.}(1999)]{Chengalur+etal+1999}
{Chengalur}, J.~N., {de Bruyn}, A.~G., \& {Narasimha}, D. 1999, \aap, 343, L79

\bibitem[{Cotten} \& {Magnani}(2013)]{Cotten+Magnani+2013}
{Cotten}, D.~L., \& {Magnani}, L. 2013, \mnras, 436, 1152

\bibitem[{Cotten} {et~al.}(2012)]{Cotten+etal+2012}
{Cotten}, D.~L., {Magnani}, L., {Wennerstrom}, E.~A., {Douglas}, K.~A., \&
  {Onello}, J.~S. 2012, \aj, 144, 163

\bibitem[{Curran} {et~al.}(2011)]{Curran+etal+2011}
{Curran}, S.~J., {Whiting}, M.~T., {Tanna}, A., {Bignell}, C., \& {Webb}, J.~K.
  2011, \mnras, 413, L86

\bibitem[{de Avillez}(2000)]{deAvillez+2000}
{de Avillez}, M.~A. 2000, \mnras, 315, 479

\bibitem[{Dickey} {et~al.}(1981)]{Dickey+etal+1981}
{Dickey}, J.~M., {Crovisier}, J., \& {Kaz\`es}, I. 1981, \aap, 98, 271

\bibitem[{Ford} {et~al.}(2010)]{Ford+etal+2010}
{Ford}, H.~A., {Lockman}, F.~J., \& {McClure-Griffiths}, N.~M. 2010, \apj, 722,
  367

\bibitem[{Grenier} {et~al.}(2005)]{Grenier+etal+2005}
{Grenier}, I.~A., {Casandjian}, J.-M., \& {Terrier}, R. 2005, Science, 307,
  1292

\bibitem[{Heiles} \& {Troland}(2003)]{Heiles+Troland+2003}
{Heiles}, C., \& {Troland}, T.~H. 2003, \apj, 586, 1067

\bibitem[{Heiles} \& {Troland}(2005)]{Heiles+Troland+2005}
{Heiles}, C., \& {Troland}, T.~H. 2005, \apj, 624, 773

\bibitem[{Kalberla} {et~al.}(2010)]{Kalberla+etal+2010}
{Kalberla}, P.~M.~W., {McClure-Griffiths}, N.~M., {Pisano}, D.~J., {et~al.}
  2010, \aap, 521, A17

\bibitem[{Komatsu} {et~al.}(2009)]{Komatsu+etal+2009}
{Komatsu}, E., {Dunkley}, J., {Nolta}, M.~R., {et~al.} 2009, \apjs, 180, 330

\bibitem[{Li} {et~al.}(2018{\natexlab{a}})]{Li+etal+2018b}
{Li}, D., {Wang}, P., {Qian}, L., {et~al.} 2018{\natexlab{a}}, IEEE Microwave
  Magazine, 19, 112

\bibitem[{Li} {et~al.}(2018{\natexlab{b}})]{Li+etal+2018a}
{Li}, D., {Tang}, N., {Nguyen}, H., {et~al.} 2018{\natexlab{b}}, \apjs, 235, 1

\bibitem[{Liszt}(1994)]{Liszt+1994}
{Liszt}, H. 1994, \apj, 429, 638

\bibitem[{Liszt} \& {Lucas}(1996)]{Liszt+Lucas+1996}
{Liszt}, H., \& {Lucas}, R. 1996, \aap, 314, 917

\bibitem[{Liszt} \& {Lucas}(2004)]{Liszt+Lucas+2004}
{Liszt}, H., \& {Lucas}, R. 2004, \aap, 428, 445

\bibitem[{Liszt} \& {Pety}(2012)]{Liszt+Pety+2012}
{Liszt}, H.~S., \& {Pety}, J. 2012, \aap, 541, A58

\bibitem[{Liszt} {et~al.}(2010)]{Liszt+etal+2010}
{Liszt}, H.~S., {Pety}, J., \& {Lucas}, R. 2010, \aap, 518, A45

\bibitem[{Liszt} \& {Wilson}(1993)]{Liszt+Wilson+1993}
{Liszt}, H.~S., \& {Wilson}, R.~W. 1993, \apj, 403, 663

\bibitem[{Lucas} \& {Liszt}(1996)]{Lucas+Liszt+1996}
{Lucas}, R., \& {Liszt}, H. 1996, \aap, 307, 237

\bibitem[{Mac Low} {et~al.}(1989)]{MacLow+etal+1989}
{Mac Low}, M.-M., {McCray}, R., \& {Norman}, M.~L. 1989, \apj, 337, 141

\bibitem[{Maller} \& {Bullock}(2004)]{Maller+Bullock+2004}
{Maller}, A.~H., \& {Bullock}, J.~S. 2004, \mnras, 355, 694

\bibitem[{McClure-Griffiths} {et~al.}(2003)]{McClure-Griffiths+etal+2003}
{McClure-Griffiths}, N.~M., {Dickey}, J.~M., {Gaensler}, B.~M., \& {Green},
  A.~J. 2003, \apj, 594, 833

\bibitem[{McClure-Griffiths} {et~al.}(2006)]{McClure-Griffiths+etal+2006}
{McClure-Griffiths}, N.~M., {Ford}, A., {Pisano}, D.~J., {et~al.} 2006, \apj,
  638, 196

\bibitem[{McClure-Griffiths} {et~al.}(2009)]{McClure-Griffiths+etal+2009}
{McClure-Griffiths}, N.~M., {Pisano}, D.~J., {Calabretta}, M.~R., {et~al.}
  2009, \apjs, 181, 398

\bibitem[{Meyer} {et~al.}(2012)]{Meyer+etal+2012}
{Meyer}, D.~M., {Lauroesch}, J.~T., {Peek}, J.~E.~G., \& {Heiles}, C. 2012,
  \apj, 752, 119

\bibitem[{Nidever} {et~al.}(2010)]{Nidever+etal+2010}
{Nidever}, D.~L., {Majewski}, S.~R., {Butler Burton}, W., \& {Nigra}, L. 2010,
  \apj, 723, 1618

\bibitem[{Peek} {et~al.}(2011{\natexlab{a}})]{Peek+etal+2011b}
{Peek}, J.~E.~G., {Heiles}, C., {Peek}, K.~M.~G., {Meyer}, D.~M., \&
  {Lauroesch}, J.~T. 2011{\natexlab{a}}, \apj, 735, 129

\bibitem[{Peek} {et~al.}(2007)]{Peek+etal+2007}
{Peek}, J.~E.~G., {Putman}, M.~E., {McKee}, C.~F., {Heiles}, C., \&
  {Stanimirovi{\'c}}, S. 2007, \apj, 656, 907

\bibitem[{Peek} {et~al.}(2011{\natexlab{b}})]{Peek+etal+2011a}
{Peek}, J.~E.~G., {Heiles}, C., {Douglas}, K.~A., {et~al.} 2011{\natexlab{b}},
  \apjs, 194, 20

\bibitem[{Putman}(2006)]{Putman+2006}
{Putman}, M.~E. 2006, \apj, 645, 1164

\bibitem[{Putman} {et~al.}(2003)]{Putman+etal+2003}
{Putman}, M.~E., {Staveley-Smith}, L., {Freeman}, K.~C., {Gibson}, B.~K., \&
  {Barnes}, D.~G. 2003, \apj, 586, 170

\bibitem[{Quilis} \& {Moore}(2001)]{Quilis+Moore+2001}
{Quilis}, V., \& {Moore}, B. 2001, \apjl, 555, L95

\bibitem[{Remy} {et~al.}(2018)]{Remy+etal+2018}
{Remy}, Q., {Grenier}, I.~A., {Marshall}, D.~J., \& {Casandjian}, J.~M. 2018,
  \aap, 611, A51

\bibitem[{Shapiro} \& {Field}(1976)]{Shapiro+Field+1976}
{Shapiro}, P.~R., \& {Field}, G.~B. 1976, \apj, 205, 762

\bibitem[{Sommer-Larsen} {et~al.}(2003)]{Sommer-Larsen+etal+2003}
{Sommer-Larsen}, J., {G{\"o}tz}, M., \& {Portinari}, L. 2003, \apj, 596, 47

\bibitem[{Stanimirovi{\'c}} {et~al.}(2008)]{Stanimirovic+etal+2008}
{Stanimirovi{\'c}}, S., {Hoffman}, S., {Heiles}, C., {et~al.} 2008, \apj, 680,
  276

\bibitem[{Wannier} {et~al.}(1993)]{Wannier+etal+1993}
{Wannier}, P.~G., {Andersson}, B.-G., {Federman}, S.~R., {et~al.} 1993, \apj,
  407, 163

\bibitem[{Wolfe} {et~al.}(2013)]{Wolfe+etal+2013}
{Wolfe}, S.~A., {Pisano}, D.~J., {Lockman}, F.~J., {McGaugh}, S.~S., \&
  {Shaya}, E.~J. 2013, \nat, 497, 224

\bibitem[{Xu} \& {Li}(2016)]{Xu+Li+2016}
{Xu}, D., \& {Li}, D. 2016, \apj, 833, 90

\bibitem[{Xu} {et~al.}(2016)]{Xu+etal+2016}
{Xu}, D., {Li}, D., {Yue}, N., \& {Goldsmith}, P.~F. 2016, \apj, 819, 22

\end{thebibliography}

\end{document}